\documentclass[11pt]{article}
\usepackage{graphicx}
\DeclareGraphicsExtensions{.eps}
\begin{document}

\title{Super-giant magnetoresistance at room-temperature \\
in copper nanowires  
due to magnetic field modulation \\ of potential barrier heights at nanowire-contact interfaces}
\author{Md. I. Hossain$^1$, M. Maksud$^2$, N. K. R. Palapati$^2$, A. Subramanian$^2$, \\
J. Atulasimha$^2$ and S. Bandyopadhyay$^{1}$ \\
\\$^1$Department of Electrical and Computer Engineering, Virginia Commonwealth University, \\ Richmond, Virginia 23284, USA \\
\\$^2$Department of Mechanical and Nuclear Engineering, Virginia Commonwealth University, \\ Richmond, Virginia 23284, USA}

\maketitle

\begin{abstract}
We have observed a super-giant ($\sim$10,000,000\%) negative magnetoresistance at 39 mT field in Cu nanowires contacted with Au 
contact pads. In these nanowires, potential
barriers form at the two Cu/Au interfaces because of Cu oxidation that results in an ultrathin 
copper oxide layer forming between Cu and Au. Current flows when electrons tunnel through, 
and/or thermionically emit 
over, these barriers. A magnetic field applied transverse to the direction of current flow 
along the wire 
deflects electrons toward one edge of the wire because of the Lorentz force, causing electron accumulation at that edge and depletion
at the other. This lowers the potential barrier at the accumulated edge and raises it at the depleted edge, causing
a super-giant magnetoresistance at room temperature. 
\end{abstract}


\vskip 0.2in

\section{Introduction}

Nanoscale magnetic field sensors have applications ranging from read heads for magnetically stored data to 
magnetic resonance imaging. Most read head sensors available today employ magnetic materials and rely on the phenomenon of 
giant magneto-resistance \cite{parkin1} or tunneling magnetoresistance in magnetic tunnel junctions \cite{parkin2}. Here, we report a magnetic field 
sesnor that does not require magnetic materials and exhibits a super-giant magnetoresistance that is several orders of 
magnitude larger than what is typically found in current nanoscale sensors.

\section{Magnetoresistance due to magnetic field modulation of the potential barrier heights at a nanowire/contact interface}

Consider a parallel array of Cu nanowires between two Au contacts as shown in Fig. 1(a). Since Cu oxidizes in the 
ambient, the nanowires will have an ultrathin Cu$_2$O or CuO coating, which will be interposed between the Au contacts 
and the Cu conductor. The bandgap of this coating is between 1.6 and 2.54 eV \cite{mugwanga}. Hence, it will result
in a potential barrier of several kT between the Cu and Au (kT is the 
room temperature thermal energy). 

\begin{figure}[!ht]
\begin{center}
\includegraphics[width=3.4in]{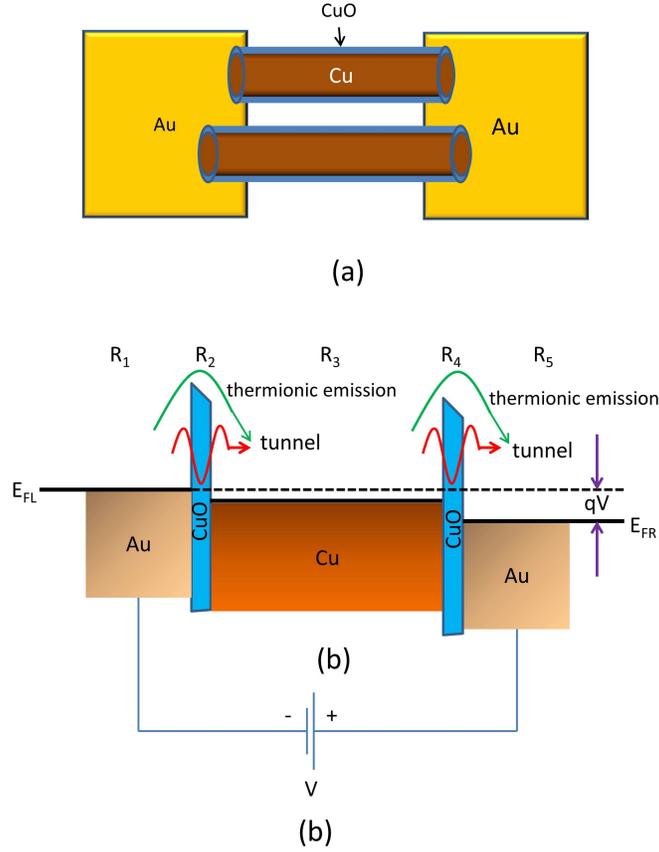}
\caption{(a) An array of Cu wires with a thin oxide coating captured between two Au contact pads. (b) The potential 
energy diagram in the direction of current flow under an applied bias $V$. The Fermi levels in the left and right
contacts are denoted by
$E_{FL}$ and $E_{FR}$, respectively, while $q$ is the electron charge. The resistance of the structure can be thought of
as being composed of five resistors in series -- the resistance $R_1$ due to the left Au contact, the resistance 
$R_2$ due to the left tunnel barrier, the resistance $R_3$ due to the Cu wire, the resistance $R_4$ due to the right
tunnel barrier and the resistance $R_5$ due to the right Au contact. The dominant resistances are $R_2$ and $R_4$.
We have arbitrarily shown only two nanowires; the effect described does not depend on the number of nanowires.}
\end{center}
\end{figure}

When a potential difference $V$ is imposed between the contacts
to induce current flow, the energy band diagram 
in the direction
of current flow  (along the wire) will appear
as shown in Fig. 1(b). Current flows  by electrons either
tunneling through, or thermionically emitting over, the barriers, or by both mechanisms. The magnitude of
the resistance will depend exponentially on the barrier heights (for thermionic emission) \cite{sze} or the square-root of 
the barrier heights (for tunneling) \cite{sze}. Any modulation of the barrier heights will therefore cause a large 
change in a wire's resistance.

\begin{figure}[!ht]
\begin{center}
\includegraphics[width=3.4in]{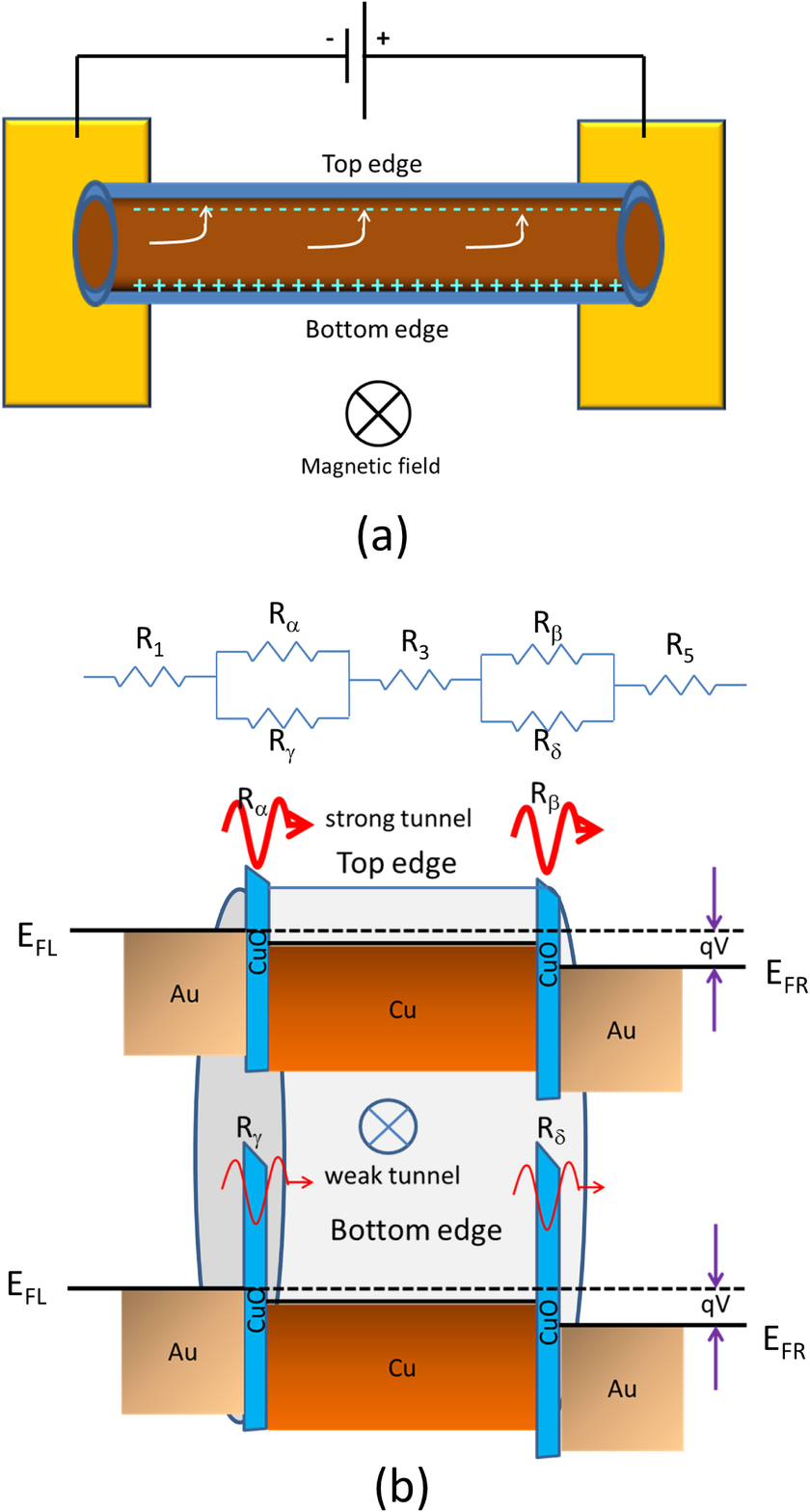}
\caption{(a) When a magnetic field is applied transverse to the direction of current flow, the resulting 
Lorentz 
force deflects electrons towards one edge, causing accumulation at one edge (in this case the top edge) and depletion
at the opposite edge (in this case the bottom edge). (b) Accumulation shifts the Fermi level upwards with respect to 
the 
conduction band edge in a metal while depletion shifts it downwards. Therefore the potential barriers become shorter at the 
accumulated 
(top) edge and taller at the depeleted (bottom edge). Electrons tunnel through (or thermionically emit over) 
the barriers much more easily
at the top edge than at the bottom edge,
making the resistances due to the barriers vastly different at the top and bottom. The top inset shows the various resistances
that contribute to the total resistance of a sample measured between the two contact pads. The dominant resistances are 
due to the tunneling/thermionic-emission barriers at the two contacts, i.e. $R_{\alpha}$, $R_{\beta}$, $R_{\gamma}$
and $R_{\delta}$ dominate and the other resistances are negligible in comparison.}
\end{center}
\end{figure}

Consider now the situation when a magnetic field is applied transverse to the direction of current flow. The resulting 
Lorentz force will divert electrons toward one edge of the wire, resulting in electron accumulation at that edge and 
depletion at the opposite edge as shown in Fig. 2(a). This is the same phenomenon that causes the classical Hall
effect. The accumulation of electrons at one edge will pull the barrier down at that edge, whereas 
the depletion of electrons at the other edge will raise the barrier up
at that edge \cite{pierret}. The sum of the amounts by which the barrier is lowered at one edge and 
raised at the other is roughly the electronic charge times the Hall voltage that appears between the 
two edges. Consequently, the energy band diagram at the two edges will look 
like in Fig. 2(b). Electrons
can tunnel (or thermionically emit) much more easily at the top edge than at the bottom edge.

As shown in Fig. 1(b), the resistance of the structure can be thought of
as being composed of five resistors in series -- the resistance $R_1$ due to the left Au contact, the resistance 
$R_2$ due to the left barriers (at the accumulated and depleted edges), the resistance $R_3$ due to the Cu wire, the resistance $R_4$ due to the right
barriers, and the resistance $R_5$ due to the right Au contact. The dominant resistances are $R_2$ and $R_4$ (since 
the other resistances are of metallic structures)
and these two are modulated by the magnetic field. 

Let us carry out a qualitative analysis of how the resistance between the 
two contacts (at low applied voltage) can change in a magnetic field due 
to  barrier height modulation at the top and bottom edges caused by the 
deflection of the electron trajectories. 
Because the analysis is ``qualitative'', 
we will view each of the resistances $R_2$ and $R_4$ as being roughly
due to {\it two} parallel paths -- one along the top edge and the other along the bottom. We will call the resistances 
of the two barriers at the top edge $R_{\alpha}$ and $R_{\beta}$, while calling the resistances of the two 
barriers at the bottom edge $R_{\gamma}$ and $R_{\delta}$ (see Fig. 2(b)). 
The resistance of the structure can therefore be written as
\begin{eqnarray}
R & = & 
R_1 + \left ( R_{\alpha} \parallel R_{\gamma} \right )  + R_3 + \left ( R_{\beta} \parallel R_{\delta} \right ) + R_5 \nonumber \\
& \approx &  \left ( R_{\alpha} \parallel R_{\gamma} \right ) + \left ( R_{\beta} \parallel R_{\delta} \right ) \nonumber \\
& = & {{1}\over{1/R_{\alpha} + 1/R_{\gamma}}} + {{1}\over{1/R_{\beta} + 1/R_{\delta}}} .
\end{eqnarray}

The resistancs $R_{\alpha}, R_{\beta}, R_{\gamma}$ and $R_{\delta}$ are determined by tunneling and/or
thermionic emission. Let us consider the tunneling case first.
Within the WKB approximation, the tunneling transmission probability of an electron impinging on a barrier with 
energy $E$
is \cite{sze}
\begin{equation}
|T(E)| = \exp \left [ - \int_{x1}^{x2} dx \sqrt{2m^* (E_c(x) - E)/\hbar^2} \right ] ,
\end{equation}
where the $x$-axis is in the direction of current flow, the tunnel barrier extends from $x = x_1$ to $x = x_2$,
and $E_c(x)$ is the spatially varying potential energy of the tunnel barrier. Similarly,
within the Richardson model, the thermionic emission probability is inversely proportional to the exponential of 
the (spatially averaged) barrier height 
at any given temperature \cite{sze}. Here, we will analyze the tunneling case. The reader can easily replicate
the analysis for the thermionic emission case. 

In the tunneling case, the resistance of any barrier is inversely proportional to the tunneling probability. Hence,
\begin{equation}
R_{m} \propto e^{\sqrt{\Delta_{m}/\Delta}} ,
\end{equation}
where $R_m$ is the resistance and $\Delta_m$ is the (spatially averaged) height of the $m$-th barrier, while $\Delta$ is a constant.

Thus, we obtain that in the absence of any magnetic field, the resistance is [from Equations (1) and (3)]
\begin{eqnarray}
R_{B = 0} & \propto &  \left [ {{1}\over{e^{-\sqrt{\Delta_{\alpha}/\Delta}} +
e^{-\sqrt{\Delta_{\gamma}/\Delta}}}} + {{1}\over{e^{-\sqrt{\Delta_{\beta}/\Delta}} +
e^{-\sqrt{\Delta_{\delta}/\Delta}}}} \right ] \nonumber \\
& \propto & \left [ e^{\sqrt{\Delta_{1}/\Delta}}{{1}\over{cosh (\Omega_1)}} + 
e^{\sqrt{\Delta_{2}/\Delta}}{{1}\over{cosh (\Omega_2)}} \right ], \nonumber \\
\end{eqnarray}
where  $\Delta_1 = \left ( \Delta_{\alpha} + \Delta_{\gamma} \right )/2$,
$\Delta_2 = \left ( \Delta_{\beta} + \Delta_{\delta} \right )/2$,
$\Omega_1 \approx \left ( \Delta_{\alpha} - \Delta_{\gamma} \right )/\left ( 4 \sqrt{\Delta_1 \Delta} \right )$,
and $\Omega_2 \approx \left ( \Delta_{\beta} - \Delta_{\delta} \right )/\left (4 \sqrt{\Delta_2 \Delta}
\right )$.

In the presence of the magnetic field, the tunnel barrier heights decrease by $\epsilon_{\alpha}$ 
and $\epsilon_{\beta}$ at the top edge, while 
increasing by $\epsilon_{\gamma}$ and $\epsilon_{\delta}$ at the bottom edge ($\epsilon_{\alpha},
\epsilon_{\beta}, \epsilon_{\gamma}, \epsilon_{\delta} > 0$). Therefore,
\begin{equation}
R_{B \neq 0}  \propto  \left [ e^{\sqrt{\Delta_1^{'}/\Delta}}{{1}\over{cosh (\Omega_1^{'})}} 
+ 
e^{\sqrt{\Delta_2^{'}/\Delta}}{{1}\over{cosh (\Omega_2^{'})}} \right ], 
\end{equation}
where
\begin{eqnarray}
\Delta_1^{'} & = & \Delta_1 + \left ( \epsilon_{\gamma} - \epsilon_{\alpha} \right ) \nonumber \\
\Delta_2^{'} & = & \Delta_2 + \left ( \epsilon_{\delta} - \epsilon_{\beta} \right ) \nonumber \\
\Omega_1^{'} & = & \Omega_1 - \left ( \epsilon_{\gamma} + \epsilon_{\alpha} \right )/\left ( 4 \sqrt{\Delta_1^{'} \Delta} \right )
\nonumber \\
& = & \Omega_1 - \omega_1 \nonumber \\
\Omega_2^{'} & = & \Omega_2 - \left ( \epsilon_{\delta} + \epsilon_{\beta} \right )/\left ( 4 \sqrt{\Delta_2^{'} \Delta} \right )
\nonumber \\
& = & \Omega_2 - \omega_2.
\end{eqnarray}
The quantities $\omega_1$ and $\omega_2$ depend on the magnetic field and are positive. Although 
$\epsilon_{\alpha} + \epsilon_{\gamma} \approx \epsilon_{\beta} + \epsilon_{\delta} \approx q V_H$, where $q$ is
the electronic charge and $V_H$ the Hall voltage that developes between the two edges, we are
unable to estimate it quantitatively since the Hall cofficient in the copper nanowires is unknown. That quantity 
depends on the free electron concentration in the nanowires, which can be several orders of magnitude different from
the bulk value owing to defects and surface states that trap electrons.

From Equations (4) - (6), we obtain that 
\begin{eqnarray}
R_{B \neq 0} - R_{B = 0} & \propto &   e^{\sqrt{\Delta_{1}^{'}/\Delta}}{{1}\over{cosh (\Omega_1^{'})}}  \nonumber \\
&&  - e^{\sqrt{\Delta_{1}/\Delta}}{{1}\over{cosh (\Omega_1)}} 
 + 
e^{\sqrt{\Delta_2^{'}/\Delta}}{{1}\over{cosh (\Omega_2^{'})}} \nonumber \\
&&  - 
e^{\sqrt{\Delta_2^{'}/\Delta}}{{1}\over{cosh (\Omega_2^{'})}} \nonumber \\
\end{eqnarray}

It is interesting to note that if the shrinkage in the barrier height at the top edge is
roughly equal to the expansion in the barrier height at the bottom edge, i.e. $\epsilon_{\alpha}
\approx \epsilon_{\gamma}$ and $\epsilon_{\beta} \approx \epsilon_{\delta}$, then $\Delta_1 \approx \Delta_1^{'}$
and $\Delta_2 \approx \Delta_2^{'}$. In that case,
\begin{eqnarray}
R_{B \neq 0} - R_{B = 0} & \propto &  e^{\sqrt{\Delta_{1}/\Delta}} \left [{{1}\over{cosh (\Omega_1 - \omega_1)}} 
- {{1}\over{cosh (\Omega_1)}}\right ]  \nonumber \\
&& + e^{\sqrt{\Delta_{2}/\Delta}} \left [{{1}\over{cosh (\Omega_2 - \omega_2)}} 
- {{1}\over{cosh (\Omega_2)}}\right ].  \nonumber \\
\end{eqnarray}
It is clear from the above expression that if in the absence of any magnetic field,
 the top and bottom barriers are similar in height, i.e. $\Omega_1 \approx \Omega_2 \approx 0$, 
 then $R_{B \neq 0} - R_{B = 0} < 0$. 
 In other words, the resistance of the structure will decrease in a magnetic field resulting in {\it
negative} magnetoresistance.

A similar analysis (algebraically easier) can be carried out for the case of thermionic emission, and the conclusion 
will be similar.

\section{Fabrication}

\begin{figure}[!ht]
\begin{center}
\includegraphics[width=3.4in]{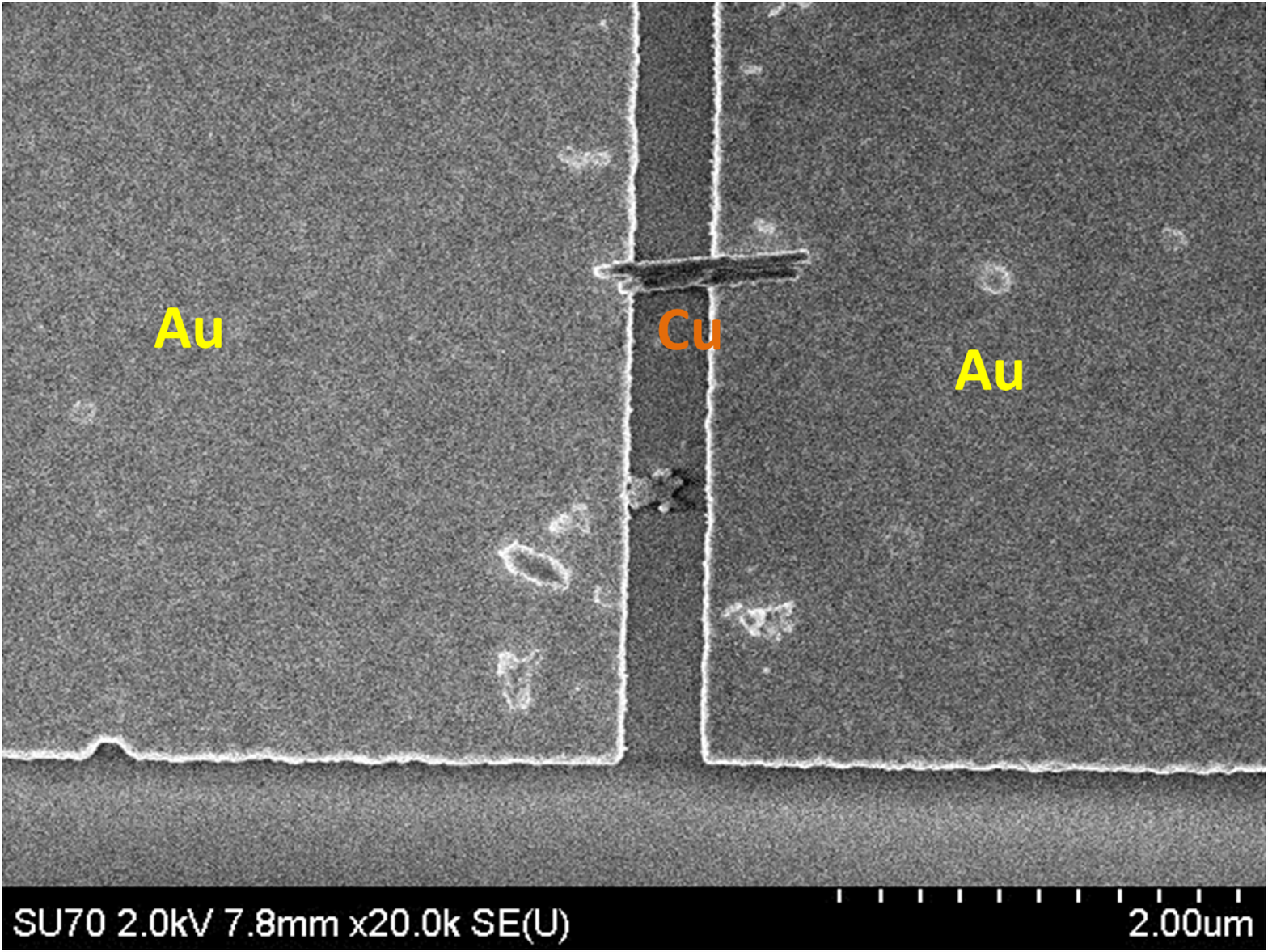}
\caption{Scanning electron micrograph of two Cu nanowires captured between two Au contact pads. }
\end{center}
\end{figure}

We fabricated 50-nm diameter Cu nanowires by depositing Cu selectively within 50-nm diameter
pores of anodic alumina films prepared by anodizing aluminum foils in 0.3M oxalic acid \cite{michalski}. 
The thin alumina layer that forms at the bottom of the pores during the anodization process impedes dc current flow along the pores; it 
was removed by soaking the alumina films in phosphoric acid for 45 - 60 minutes \cite{saumil}.
Next, Cu was selectively electrodeposited within the nanopores from a solution of CuSO$_4$ \cite{JEM} using a dc current density 
of 1 mA/cm$^2$. During this step, the sample was immersed in the CuSO$_4$
solution and current was passed through the solution using the aluminum foil as the cathode 
and a counter electrode as the anode. The Cu$^{++}$ cations flowed into the pores from the solution because the pores 
offered the 
least resistance paths for the dc current. Thus,
Cu nanowires of 50 nm diameter formed within the pores of the alumina film. The alumina host was dissolved out
in hot chromic/phosphoric acid, leaving the Cu nanowires standing vertically on the aluminum foil. 
The samples were then
ultrasonicated in ethanol to release the wires from the aluminum foil, forming a suspension of Cu nanowires
(50 nm diameter and varying length; average length $\sim$ 1 $\mu$m) in ethanol.

The Cu nanowires in the ethanol suspension were captured across Au contact pads on silicon chips using dielectrophoresis 
\cite{arun1,arun2,arun3}. This involved 
the following steps. Electron beam lithography was used to create patterns in poly-methyl methacrylate (PMMA) based 
polymeric resists, which were spin-coated on silicon chips. An 85-nm thick Au film (with a Cr adhesion layer) 
was deposited on the resist-patterned surface and metal lift-off was performed to create spatially separated 
nanoelectrode pairs on the chips. Next, the nanowire containing ethanol suspension was pipetted on to the surface of 
the silicon chip patterned with the
Au nanoelectrodes, while applying an ac bias across the nanoelectrodes. This polarized the Cu nanowires in the 
suspension and exerted a dielectrophoretic (DEP) force that caused  them to be captured across the nanoelectrodes. 
By tuning the excitation voltage (4V, peak-to-peak) and its frequency (1 kHz) over a given deposition period 
(4 minutes), the DEP region of influence was controlled to extend over suspension volumes that present single or a 
few nanowires \cite{arun3}. This resulted in an array of devices on the chip with a single or a few nanowires 
assembled across 
the Au nanoelectrodes. From these assembled device arrays, single (or few) nanowire assembly locations were selected 
for further 
testing. A scanning electron microscope image of a 
representative device with two
assembled nanowires is shown in Fig. 3.

\section{Results}

\begin{figure}[!ht]
\begin{center}
\includegraphics[width=3.4in]{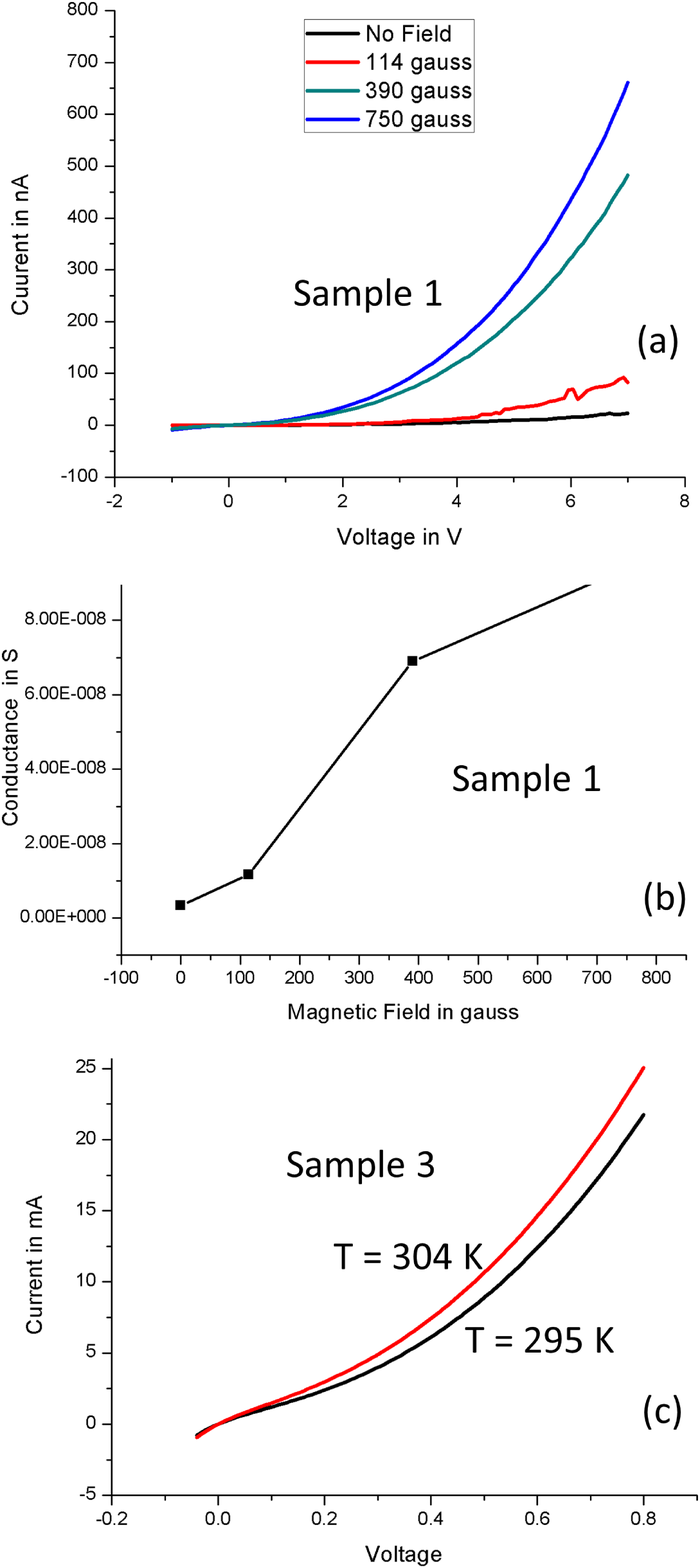}
\caption{(a) Current versus voltage characteristic of a sample plotted for zero 
and three different magnetic field strengths. (b) Magnetoconductance of the sample (conductance 
versus magnetic field) at a fixed voltage of 7 V. (c) Current-voltage characteristic of a sample measured
at two slightly different temperatures.}
\end{center}
\end{figure}

\begin{figure}[!ht]
\begin{center}
\includegraphics[width=3.4in]{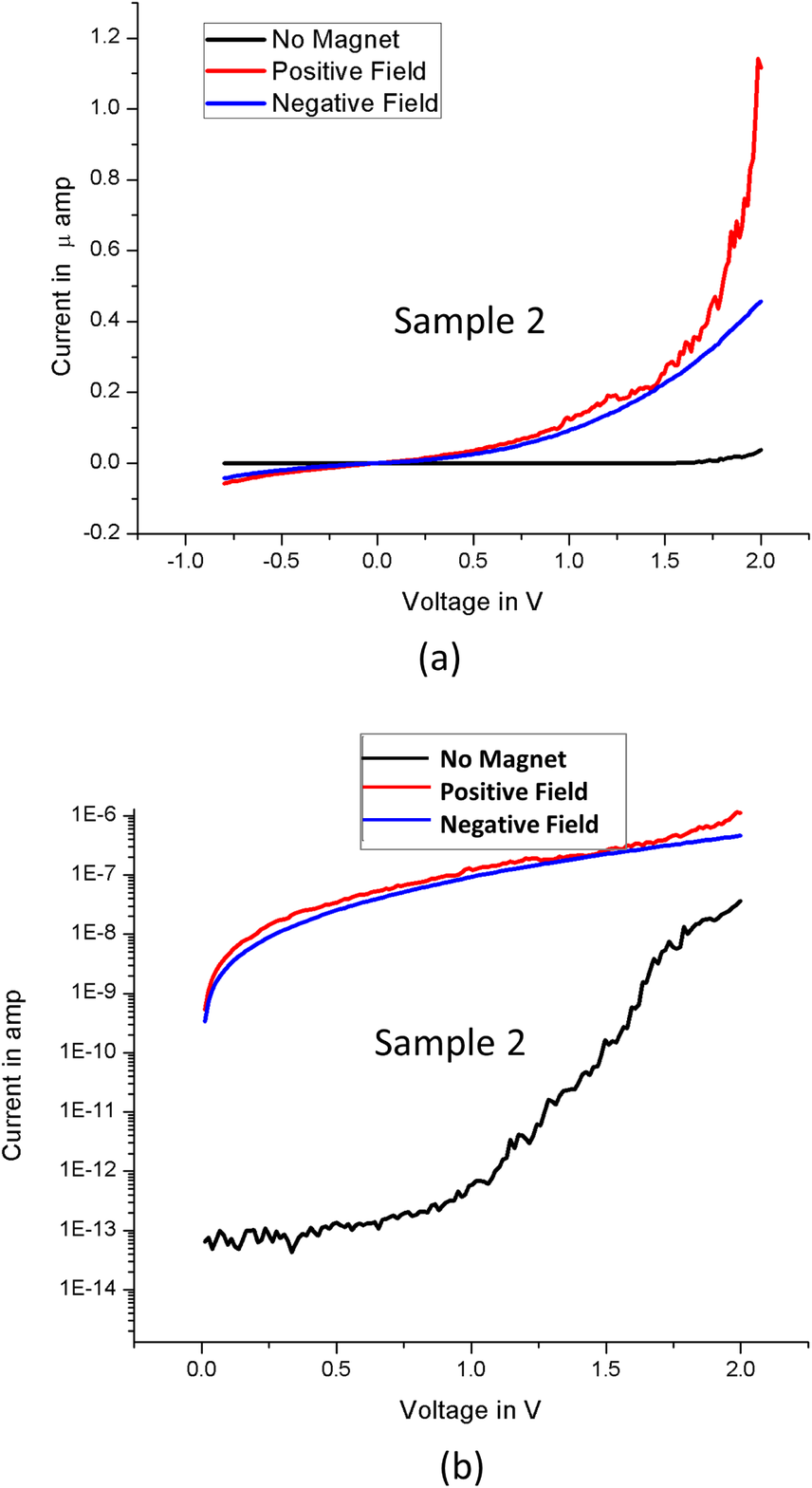}
\caption{(a) Current versus voltage characteristics of a sample at zero magnetic field 
and at two opposite directions of a magnetic field of 39 mT. (b) The same 
current-voltage characteristics plotted in a log-linear scale.}
\end{center}
\end{figure}

The current voltage characteristics of the nanowires were measured at room temperature with a HP 4156B semiconductor 
parameter analyzer under different magnetic fields. The plot for one sample is shown in Fig. 4 (a). These plots are 
repeatable from run to run and do not show any instability,

The characteristics in Fig. 4(a) are 
expectedly non-linear because of the tunneling through, and/or thermionic emission over, the barriers. The current at 
a given voltage increases with increasing magnetic field, resulting in a negative magnetoresistance
(or positive magnetoconductance). The conductance at a fixed voltage of 7 V is plotted as a function of magnetic field in
Fig. 4(b). In our set up,  the maximum magnetic field we can apply is 75 mT. 

In Fig. 4(c), we show the current-voltage characteristic of a third sample at zero magnetic field measured
at two different temperatures. Increased temperature increases the conductance, despite the fact that phonon scattering 
in copper should incerase when the sample is heated.
This increase happens because transport
occurs via thermionic emission over a barrier and the emission rate increases with temperature. The increase is 
however relatively 
modest which tells us that thermionic emission is not the only modality of transport; there is also significant 
tunneling through the barrier.

In Fig. 5(a),
we show the current-voltage plots for a fixed magnetic field strength of 39 mT, but for two anti-parallel directions of the field.
 There 
is a difference between the two cases since the electrons are deflected to opposite edges and the 
transport characteristics of the two edges are, expectedly, somewhat different. In this case, we have also plotted the 
current-voltage 
characteristic in log-linear scale in Fig. 5(b) to show that the magnetoconductance is astoundingly large -- the resistance can change by 
over five orders of magnitude in a magnetic field of 39 mT under an applied voltage of 250 mV (Fig. 5), 
resulting in a super-giant magnetoresistance of 10,000,000\%
at that magnetic field. The low voltage makes it an extremely low power 
device -- the maximum power dissipation at that voltage level is $\sim$5 nW.

One disconcerting feature is that there is significant variability in the absolute conductance of the samples, as
is evident from the current-voltage characteristics of samples 1, 2 and 3. This is obviously due to the fact that the 
copper oxide film cannot be well controlled and the current is inversely proportional to the exponential of 
the barrier height for thermionic emission (square-root of barrier height for tunneling). In the tunneling case,
the current is also inversely proportional to the exponential of the barrier width, or the oxide film thickness. 
Because of this variability, some samples do not show the effect while others do. Better control
may be achieved by growing the oxide layer with atomic layer deposition, instead of relying on atmospheric oxidation.
This is, however, a yield-related concern that does not impute the physics of the effect.

\section{Conclusions}

In conclusion, we have shown that a giant magnetoconductance can result in a metal nanowire from magnetic field 
modulation of potential barrier heights at 
metal-contact interfaces.  The field deflects 
electron trajectories toward one edge of the metal nanowire which causes the local electron concentration
at that edge to rise while the concentration at the other edge falls.
This lowers the potential barrier height at the first edge and raises it at the second. The effect of 
this is to alter the nanowire's conductance and the change in the conductance can be super-giant. 
This effect is different from the traditional giant magnetoresistance effect accruing from spin-dependent scattering in magnetic multilayers
\cite{fert,grunberg}, or tunneling magnetoresistance in magnetic tunnel junctions
\cite{ikeda}, or tunneling of electrons (injected by a ferromagnetic contact) between Landau levels in adjacent graphene 
layers \cite{sandipan}.
It is much larger and can be used in all the
traditional applications of giant magnetoresistance, such as in magnetic read heads and magnetic field sensors.

\section*{Acknowledgements}

This work was supported by the US National Science Foundation under grant CMMI-1301013.


\end{document}